\def\plotfiddle#1#2#3#4#5#6#7{\centering \leavevmode
\vbox to#2{\rule{0pt}{#2}}
\includegraphics{#1}}

\def \kmsmpc {\>{\rm km}\,{\rm s}^{-1}\,{\rm Mpc}^{-1}}
\def \etal {et~al.~}
\def \Yd   {\ifmmode \Upsilon_{\rm d} \else $\Upsilon_{\rm d}$ \fi}
\def \YdR  {\ifmmode \Upsilon_{\rm d}^R \else $\Upsilon_{\rm d}^R$ \fi}
\def \YI  {\ifmmode \Upsilon_I \else $\Upsilon_I$ \fi}
\def \LI  {\ifmmode L_I \else $L_I$ \fi}
\def \Rd   {\ifmmode R_{\rm d} \else $R_{\rm d}$ \fi}
\def \rs   {\ifmmode r_{\rm s} \else $r_{\rm s}$ \fi}
\def \rrm2  {\ifmmode r_{-2} \else $r_{-2}$ \fi}
\def \ccm2  {\ifmmode c_{-2} \else $c_{-2}$ \fi}
\def \cvir {\ifmmode c_{\rm vir} \else $c_{\rm vir}$ \fi }
\def \Rvir {\ifmmode R_{\rm vir} \else $R_{\rm vir}$ \fi}
\def \Vvir {\ifmmode V_{\rm vir} \else $V_{\rm vir}$ \fi}
\def \Mvir {\ifmmode M_{\rm vir} \else $M_{\rm vir}$ \fi}
\def \Msun {\ifmmode M_{\odot} \else $M_{\odot}$ \fi}
\def \Deltavir {\ifmmode \Delta_{\rm vir} \else $\Delta_{\rm vir}$ \fi}
\def \OmegaM {\ifmmode \Omega_{\rm M} \else $\Omega_{\rm M}$ \fi}
\def \lam {\ifmmode \lambda \else $\lambda$ \fi}
\def \lamp {\ifmmode \lambda^{\prime} \else $\lambda^{\prime}$ \fi}
\def \lambar {\ifmmode \bar{\lambda} \else $\bar{\lambda}$ \fi}
\def \lampbar{\ifmmode \bar{\lambda^{\prime}} \else $\bar{\lambda^{\prime}}$\fi}
\def \siglam {\ifmmode \sigma_{\lambda} \else $\sigma_{\lambda}$ \fi}
\def \siglamp {\ifmmode \sigma_{\lambda^{\prime}} \else 
      $\sigma_{\lambda^{\prime}}$ \fi}
\def \sigl {\sigma_{\ln\lambda}} 
\def \Sigmacrit {\ifmmode  \Sigma_{\rm crit} \else $\Sigma_{\rm crit}$ \fi}  
\def \md {\ifmmode  m_{\rm d}  \else $m_{\rm  d}$ \fi}  
\def \ms {\ifmmode m_{\rm s}  \else $m_{\rm s}$ \fi} 
\def \mdbar {\ifmmode  {\overline{m}}_{\rm d}  \else ${\overline{m}}_{\rm  d}$ \fi}  
\def \msbar {\ifmmode  \bar{m}_{\rm s}  \else $\bar{m}_{\rm  s}$ \fi}
\def \cbar {\ifmmode  \overline{c} \else $\overline{c}$ \fi}
\def  \Md {\ifmmode M_{\rm d} \else $M_{\rm  d}$ \fi} 
\def \Ms {\ifmmode  M_{\rm s} \else $M_{\rm s}$ \fi} 
\def \Mstar {\ifmmode  M_{\rm star} \else $M_{\rm star}$ \fi}
\def \Mdisc {\ifmmode M_{\rm disc}  \else $M_{\rm disc}$ \fi} 
\def \Rd {\ifmmode R_{\rm d}  \else $R_{\rm d}$ \fi} 
\def  \Rs {\ifmmode R_{\rm s} \else $R_{\rm s}$ \fi} 
\def \VIII {\ifmmode V_{3.2} \else $V_{3.2}$ \fi}  
\def \VII  {\ifmmode  V_{2.2} \else  $V_{2.2}$  \fi} 
\def  \Vobs {\ifmmode V_{\rm  obs} \else $V_{\rm obs}$ \fi}  
\def \Vdisc {\ifmmode V_{\rm disc} \else $V_{\rm disc}$ \fi}
\def \Vhalo {\ifmmode V_{\rm halo} \else $V_{\rm halo}$ \fi} 
\def \RIII {\ifmmode 3.2\Rs \else $3.2\Rs$ \fi} 
\def \RII {\ifmmode 2.2\Rs \else $2.2\Rs$ \fi} 
\def \R200 {\ifmmode R_{200} \else $R_{200}$ \fi} 
\def \v200 {\ifmmode V_{200}  \else  $V_{200}$  \fi}  
\def \M200  {\ifmmode  M_{200}  \else $M_{200}$ \fi}
\def \qd {\ifmmode q_{\rm d} \else $q_{\rm d}$ \fi} 
\def \qh {\ifmmode q_{\rm h} \else $q_{\rm h}$ \fi}
\def  \vmax  {\ifmmode  V_{\rm  max}  \else $V_{\rm  max}$  \fi}  
\def \vmaxobs{\ifmmode V_{\rm max}^{\rm  obs}\else $V_{\rm  max}^{\rm obs}$\fi} 
\def \vdisc {\ifmmode V_{\rm disc} \else $V_{\rm disc}$ \fi}
\def \vtot {\ifmmode V_{\rm tot}  \else $V_{\rm tot}$ \fi} 
\def \vcirc {\ifmmode V_{\rm circ} \else  $V_{\rm circ}$ \fi} 
\def \vrot {\ifmmode V_{\rm rot} \else $V_{\rm rot}$ \fi}
\def \kms {\ifmmode  \,\rm km\,s^{-1} \else $\,\rm km\,s^{-1}  $ \fi }
\def  \kpc {\ifmmode  {\rm kpc}  \else ${\rm  kpc}$ \fi  }  
\def \Msol {\ifmmode  \rm  M_{\odot} \else  $\rm  M_{\odot}$  \fi  } 
\def \hMsol {\ifmmode  h^{-1}\,\rm M_{\odot} \else $h^{-1}\,\rm M_{\odot}$ \fi}
\def \Msolpc2{\ifmmode\rm M_{\odot}\,pc^{-2}\else $\rm M_{\odot}\,pc^{-2}$\fi}
\def  \chisq  {\ifmmode  \chi^2   \else  $\chi^2$  \fi}  
\def  \chisqr {\ifmmode \chi^2_{\rm r} \else $\chi^2_{\rm r}$ \fi}
\def  \spose#1{\hbox  to 0pt{#1\hss}}  
\def\lta{\mathrel{\spose{\lower 3pt\hbox{$\sim$}}\raise  2.0pt\hbox{$<$}}}
\def\gta{\mathrel{\spose{\lower  3pt\hbox{$\sim$}}\raise 2.0pt\hbox{$>$}}}

\documentclass{PoS} \usepackage{epsfig}

\PoS{PoS(BDMH2004)050}

   \title{Origin of the Joint Distribution of Structural Parameters in 
          Disc Galaxies}

 \ShortTitle{Structural Correlations in Disc Galaxies}

\author{\speaker{Aaron  A.    Dutton}\\  ETH  Z\"urich,  Switzerland\\
E-mail: \email{dutton@phys.ethz.ch} }

\author{Frank C. van den  Bosch\\ ETH Z\"urich, Switzerland \\ E-mail:
        \email{vdbosch@phys.ethz.ch} }

\author{St\'ephane  Courteau\\  Queen's  University, Canada\\  E-mail:
\email{courteau@astro.queensu.ca}}

\author{Avishai  Dekel\\  The   Hebrew  University,  Israel\\  E-mail:
\email{dekel@astro.huji.ac.il}}

\abstract{  We use  a simple  model  of disc-galaxy  formation in  the
  $\Lambda$CDM  cosmology  to  simultaneously  reproduce  the  slopes,
  zero-points,  scatter  and uncorrelated  residuals  in the  observed
  velocity-luminosity  (VL)  and  radius-luminosity  (RL)  relations.  
  Observed I-band  luminosities are converted to  stellar masses using
  the  IMF-dependent  relation  between stellar  mass-to-light  ratio,
  $\Upsilon$, and  color.  The  model treats halo  concentration, spin
  parameter, and  disc mass fraction as  independent log-normal random
  variables.  Our  main conclusion is  that the VL and  RL zero-points
  and the uncorrelated residuals can only be reproduced simultaneously
  if {\it adiabatic contraction is avoided}. One or the other could be
  fixed  by  appealing  to  unrealistic values  for  $\Upsilon$,  halo
  concentration  $c$,  and  spin  parameter  $\lambda$,  but  not  all
  together.   The small  VL  scatter is  naturally  determined by  the
  predicted scatter in $c$ and in $\Upsilon$, quite independent of the
  large scatter in $\lambda$.  However,  the RL scatter, driven by the
  scatter  in $\lambda$,  can be  as low  as observed  only  if $\sigl
  \simeq 0.25$, about  half the value predicted for  CDM haloes.  This
  may indicate that  discs form in a special subset  of haloes. The VL
  slope  is  reproduced  once  star  formation  occurs  only  above  a
  threshold surface  density, and the  RL slope implies that  the disc
  mass fraction is increasing with halo mass, consistent with feedback
  effects.  A model that incorporates the above ingredients provides a
  simultaneous fit  to all the  observed features. In  particular, the
  elimination of  adiabatic contraction allows  80\% disc contribution
  to the observed rotation velocity  of bright discs at 2.2 disc scale
  lengths.   The  lack of  halo  contraction  may  indicate that  disc
  formation is not as smooth  as typically envisioned, but instead may
  involve clumpy, cold streams.}
 
\FullConference{Baryons  in  Dark Matter  Halos\\  5-9 October  2004\\
 Novigrad, Croatia}

\begin{document}

\section{Introduction}

The  correlations  between the  global  structure  parameters of  disc
galaxies,   rotation  velocity  $V$,   exponential  radius   $R$,  and
luminosity  $L$,  provide  invaluable  constraints  on  the  formation
process of these galaxies.  Although various studies have attempted to
explain the  origin of the RL  or the VL (also  known as Tully-Fisher)
relation, the  true challenge lies in finding  a self-consistent model
of  disc  formation  that  can  match both  simultaneously.   Here  we
investigate  how  standard  models  of  disc formation  fair  in  this
respect. 

We use the  observed structural parameters from an  $I$-band sample of
$\sim 1800$  bright disc galaxies  compiled by Courteau  \etal (2003).
In order to facilitate a simple comparison with theory, we convert the
observed $I$-band  luminosity into a  stellar masses $\Ms$,  using the
stellar mass-to-light ratio $\YI$ at a given color as obtained by Bell
\etal (2003) and Bell \& de Jong (2001).  The main uncertainty in this
conversion  is  the normalization  of  $\YI$,  reflecting the  unknown
stellar initial  mass function (IMF).   We adopt the IMF  advocated by
Bell  \etal  (2003) and  note  that  the  most extreme  top-heavy  IMF
realistically  considered  (the Kroupa-IMF;  Kroupa,  Tout \&  Gilmore
1993) has values of $\log(\YI)$ that  are only $0.15$ dex smaller.  We
use  as joint constraints  on our  model the  observed log  slopes and
zero-points  of the  conditional $V\vert\Ms$  and  $R\vert\Ms$ scaling
laws,  the  conditional  scatter  about  each  at  a  given  $M$,  and
especially the  cross-correlation between the  corresponding residuals
$\Delta   V$  and   $\Delta  R$   at  $\Ms$.   Throughout   we  assume
$H_0=70\kmsmpc$.

\section{Disc Formation Models}

Our  model starts  in the  spirit of  the classical  model by  Fall \&
Efstathiou (1980) and  Mo, Mao, \& White (1998,  hereafter MMW).  Each
model galaxy  is specified by  four parameters: the total  virial mass
$M_{\rm vir}$,  the halo concentration parameter $c$,  the baryon spin
parameter $\lambda$,  and the mass fraction  that ends up  in the disc
$\md$.  The dark-matter  is assumed to follow a  spherical NFW density
profile, with $c$ that  varies systematically with $M_{\rm vir}$.  The
disc, with mass  $M_{\rm d} = \md \, M_{\rm vir}$,  is assumed to have
an exponential density profile, whose scale length $R_{\rm d}$ follows
from  $\lambda$, $\md$,  and $M_{\rm  vir}$ based  on angular-momentum
conservation as outlined in MMW.   In this standard model, the halo is
assumed to  contract adiabatically under  the dissipative condensation
of the gas following Blumenthal \etal (1986).

A major novel ingredient considered  here is the possible avoidance of
the  adiabatic  contraction,  hereafter  AC  (see  \S~4  for  possible
physical motivation).  Another straightforward novelty adopted here is
splitting  the disc  into a  gaseous and  a stellar  component.  Using
Toomre's critical surface density  $\Sigma_{\rm crit}$, we compute the
stellar  mass as  the  disc  mass with  surface  density $\Sigma  \geq
\Sigma_{\rm  crit}$, mimicking a  star-formation threshold  density in
agreement with  the empirical findings of  Kennicutt (1989).  Although
this is clearly an oversimplification, it helps explaining (a) why the
gas mass fraction  is higher in lower mass  galaxies (Kannappan 2004),
(b) why the  scale length of the  cold gas is larger than  that of the
stars, and  (c) the  truncation radii of  stellar discs (e.g.  van den
Bosch 2001).

The parameters $c$, $\lambda$ and  $\md$ are assumed to be independent
random  variables, each with  a log-normal  distribution in  which the
mean could vary  systematically with $\Mvir$.  In order  to pursue the
comparison  to observations,  we construct  a large  sample  of random
realizations  of  model  galaxies.   For  a given  $\Mvir$,  the  halo
concentration  $c$  is  drawn  from a  log-normal  distribution,  with
$\cbar=  12 (  V_{\rm vir}  /  100 )  ^ {  -0.33}$ and  $\sigma_{\ln\,
c}=0.32$, based on the CDM-simulated haloes (Eke, Navarro, \& Steinmetz
2001; Bullock \etal 2001a).   The baryonic spin parameter $\lambda$ is
initially  drawn  from an  independent  log-normal distribution,  with
$\lambar=0.035$  and  $\sigma_{\ln\lambda}=0.5$,  again based  on  the
simulated properties of  CDM haloes (Bullock \etal 2001b),  but we are
forced to allow lower scatter  values (see below).  We then draw $\md$
from  an independent  log-normal distribution,  but  consider $\mdbar$
$(\Mvir)$ and $\sigma_{{\rm ln}  m_{\rm d}}$ as free parameters.  Once
we compute  the disc  mass and radius,  we derive the  {\it stellar\,}
surface  density  profile and  evaluate  its  mass $\Ms$,  exponential
radius $\Rs$, and ``observed" circular velocity $\VIII$ at $3.2\Rs$.

For a fair  comparison with the data, we  attempt to reproduce similar
sampling  in  the model  realizations.   We  first  sample with  equal
probability in $\log\Mvir$, and then  select a subsample for which the
distribution of the  derived $\VIII$ resembles that of  the data.  The
model scaling relations and scatter are obtained by linear regressions
in log space of $V$ on $M$ and $R$ on $M$.

\section{Comparison of Models and Observations} 

\begin{table}
\center
\begin{tabular}[t]{l|ll|ll}
$Y\propto X^{\rm  \;slope}$ & $\;\;\;Y\;-\;X$ & slope  & $\,Y\;-\;X$ &
slope\\ \hline \hline $I$-Band & $\Vobs-\LI$ & 0.32 & $\Rs-\LI$ & 0.34
\\ $\Mstar$  & $\Vobs-\Ms$ & 0.26 &  $\Rs-\Ms$ & 0.27 \\  \hline MMW &
$\VIII-\Md$ & 0.33  & $\Rd-\Md$ & 0.33 \\  $+c(\Mvir)$ & $\VIII-\Md$ &
0.31 & $\Rd-\Md$ & 0.36 \\ $+\Sigma_{\rm crit}$ & $\VIII-\Ms$ & 0.27 &
$\Rs-\Ms$ & 0.35  \\ $+m_d(\Mvir)$ & $\VIII-\Ms$ &  0.25 & $\Rs-\Ms$ &
0.27 \\ \hline
\end{tabular}
\caption{Slopes of observed versus model scaling relations.  
 The upper 2 rows refer to the observations, while the lower 5 rows to
 the model.  $\Vobs$ is the observed  (inclination corrected) rotation
 velocity. $\VIII$  is the model circular velocity  at $3.2\Rs$. $\LI$
 is the observed (dust corrected) $I$-band luminosity. $\Ms$ and $\Md$
 are the model  stellar and total disc mass,  respectively.  $\Rs$ and
 $\Rd$ are the corresponding disc scale lengths.}
\end{table}

\subsection{Slopes}

We start our exploration with the simplest MMW model possible. We keep
$c$, $\lambda$,  and $\md$ fixed  at 12, 0.035 and  0.05 respectively,
and  compute the  (scatter-free)  scaling relations  VM  and RM.   The
apparent  agreement  between the  predicted  slopes  and the  $I$-band
scaling relations, listed in lines 3 and 1 of Table~1 respectively, is
misleading. This comparison naively assumes  that all the disc mass is
transformed into  stars and with  a constant $\YI$.  A  more realistic
comparison of or model is with the relations listed in the second line
of  Table~1, where  $\Ms$ replaces  $\LI$ in  the data.   Clearly, the
basic MMW  model yields  steeper slopes than  observed.  The  model VM
slope gets  smaller when the proper  scaling of $c$ with  halo mass is
incorporated, and  it becomes almost  as shallow as observed  when the
star-formation threshold density is  imposed.  A simultaneous match of
the RM slope is obtained when  $\md$ is taken to scale with halo mass,
$\md  \propto \Mvir^{0.16}$,  as qualitatively  expected  from stellar
feedback effects and as inferred from rotation curves (Persic, Salucci
\& Stel 1996).

\subsection{Zero-points}

\begin{figure}
\plotfiddle{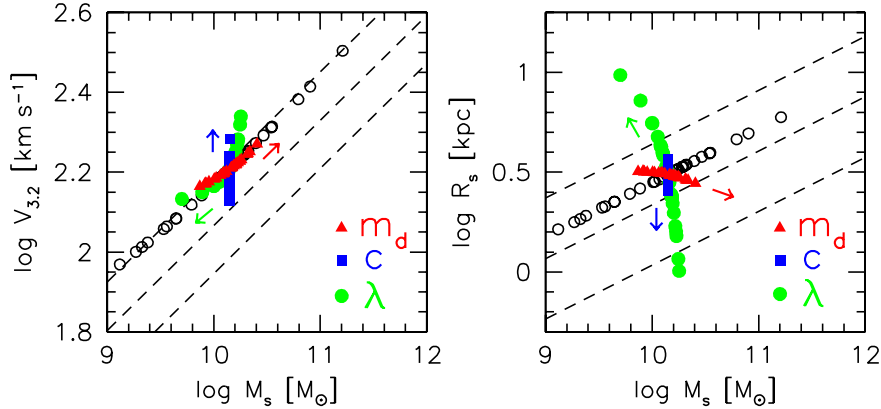}{1.8in}{0}{60}{60}{-180}{-260}
\caption{Scaling relations for the model which matches the slopes
  (last row of  Table~1, open circles), versus the  mean and 2$\sigma$
  scatter of  the observed relations  (dashed lines).  The  effects of
  2$\sigma$ variations  in $\md$, $c$  and $\lambda$ are  shown, where
  $\sigma_{\ln\,  m_{\rm  d}}=0.23$,  $\sigma_{\rm ln\,  c}=0.32$  and
  $\sigma_{\rm ln\, \lambda}=0.50$.  The arrows indicate the direction
  of increasing $\md$, $c$, and $\lambda$.}
\end{figure}
Fig.~1 demonstrates that the model  which matches the slopes in \S 3.1
suffers  from  severe  zero-point  offsets compared  to  the  observed
relations.  This could have been fixed by a $\sim 0.4$ dex decrease in
$\YI$, but realistic IMFs limit the correction to $0.15$ dex, which we
adopt  in our  models below.   Also seen  in Fig.~1  is the  impact of
varying $c$,  $\lambda$ and $\md$  over the 2-$\sigma$ range  of their
distributions.   Clearly, the  VM  zero-point is  mostly sensitive  to
$\cbar$, while the RM zero-point is most sensitive to $\lambar$ and to
a lesser extent to the other variables (the secondary dependencies are
partly due to the effect of AC).

The RM zero-point  could have been matched by  decreasing $\lambar$ or
increasing $\mdbar$, but  these changes would have left  the VM offset
unchanged or  worsen it.  Furthermore,  such changes would  worsen the
conflict with the observed, uncorrelated residuals (see \S~3.4 below),
and thus fail to provide  a simultaneous fit.  The VM zero-point could
have been  matched with  the acceptable $0.15$  dex decrease  in $\YI$
plus  a  factor of  2  decrease  in  $\cbar$, \footnote{Introducing  a
central  density  core does  not  affect  $\VIII$ significantly  (e.g.
Dutton  \etal   2003).}   but  this  requires  a   drastic  change  of
cosmological  parameters,  and  also  worsens  the  problem  with  the
observed, uncorrelated residuals.

The  above  is  a  reflection  of  a well  known  failure  of  current
semi-analytical   models  of   galaxy  formation,   being   unable  to
simultaneously  fit the  luminosity  function (and  hence the  average
virial mass-to-light  ratio) and the  TF zero-point.  A  VM zero-point
match could  be obtained if  $\Vobs \simeq \Vvir$, but  realistic halo
concentrations   in   the   standard   $\Lambda$CDM   cosmology   with
$\sigma_8=0.9$  imply  $\Vobs/\Vvir   \simeq  1.4$,  and  standard  AC
actually boosts it up to $\Vobs/\Vvir \simeq 1.6$.

We  are  led  to  the  conclusion  that  the  only  way  to  obtain  a
simultaneous match to the slopes and zero-points of both the VM and RM
relations,  within the  realistic  range of  parameter  values in  the
$\Lambda$CDM  scenario, and  with  a hope  to  match the  uncorrelated
residuals (\S 3.4), is by weakening or completely removing the effects
of AC on $\VIII$ (see Fig.~3).

\subsection{Scatter}

The  observed scatters  about the  scaling relations  are $\sigma_{\rm
  ln}(V|L_I)=0.14$ and $\sigma_{\rm ln}(R|L_I)=0.35$.  The measurement
  errors are estimated to be $\sigma_{\rm ln} \simeq 0.09$ for each of
  the  observables $V$,  $R$,  and  $\LI$.  In  addition,  there is  a
  contribution to the scatter in our evaluated $\Ms$ due to scatter in
  the  relation between  color  and  $\YI$, which  we  estimate to  be
  $\sigma_{\ln\,\YI}\gta0.23$   based   on   Bell  \etal(2003).    The
  remainder of the  observed scatter is assumed to  originate from the
  intrinsic scatter in $c$, $\lambda$, and $\md$.

Table~2  illustrates  how  each  of  these  five  sources  of  scatter
contributes  to the  total scatter  in the  two relations.   Note that
although  the   different  sources  of  scatter  are   assumed  to  be
uncorrelated, the  residuals in the derived  quantities ($\LI$, $\Rs$,
$\VIII$) are  partly correlated.  This explains why  the total scatter
is less than the quadratic sum of the scatter in each component.

As  evident  from  Fig.~1   and  Table~2,  the  scatter  in  $\lambda$
completely dominates the RM scatter.  The observed RM scatter places a
robust upper  limit of $\sigma_{\ln\lambda}  \simeq 0.25$, independent
of  the other  properties  of the  structural  correlations.  This  is
roughly  half the  value found  in cosmological  simulations,  and may
indicate that disc formation only occurs in a subset of haloes (with a
restricted range in $\lambda$).

In  the VL  scatter,  the contributions  of  observational errors  and
scatter in $\YI$ and in $c$ are all comparable, while the contribution
of  $\sigma_{\ln\lambda}=0.25$  is  smaller,  and  that  of  $\md$  is
negligible.  While it would have been possible for scatter in $\YI$ or
$c$ alone to account for all the intrinsic VL scatter, each would have
resulted  in  a strong  residual  correlation,  in  conflict with  the
observations (\S 3.4).  We therefore  conclude that {\it both} $c$ and
$\YI$ have to contribute comparably to the VL scatter.
Note  that  with  a  significantly higher  $\sigma_{\ln\lambda}$,  the
residuals  about  the VL  relation  would  have  been correlated  with
$\lambda$.  Given that the radius, and thus the surface brightness, is
a   strong  function  of   $\lambda$,  this   would  have   implied  a
surface-brightness  dependence   in  the  VL   relation,  contrary  to
observations (Courteau \& Rix 1999; hereafter CR99).

\begin{table}[h]
\center
\begin{tabular}{l|l|ccccc|cc}
$\sigma_{\ln\,\lambda}$ & Relation & $c$ & $\lambda$ & $\md$ & $\YI$ &
err & total & obs \\  \hline 0.50 & $\sigma_{\rm ln} (\Rs|\LI)$ & 0.09
& 0.56 & 0.13  & 0.07 & 0.10 & 0.67 & 0.35  \\ 0.25 & $\sigma_{\rm ln}
(\Rs|\LI)$ & & 0.32  & & & & 0.37 & \\  \hline 0.50 & $\sigma_{\rm ln}
(\VIII|\LI)$ & 0.09 & 0.08 & 0.02 &  0.07 & 0.09 & 0.16 & 0.14 \\ 0.25
& $\sigma_{\rm ln} (\VIII|\LI)$ & & 0.05 & & & & 0.13 & \\ \hline
\end{tabular}
\caption{sources of scatter in the size-luminosity and
  velocity-luminosity relations. The intrinsic sources of scatter are:
$\sigma_{\ln\,c}=0.32$;   $\sigma_{\ln\,\lambda}=0.25\,\rm  or\,0.50$;
$\sigma_{\ln\,\md}=0.23;$  $\sigma_{\ln\,\YI}=0.23$,  and $\sigma_{\ln
\rm err}=0.09$.}

\end{table}

\subsection{Residual Correlation}
\begin{figure}[t]
\plotfiddle{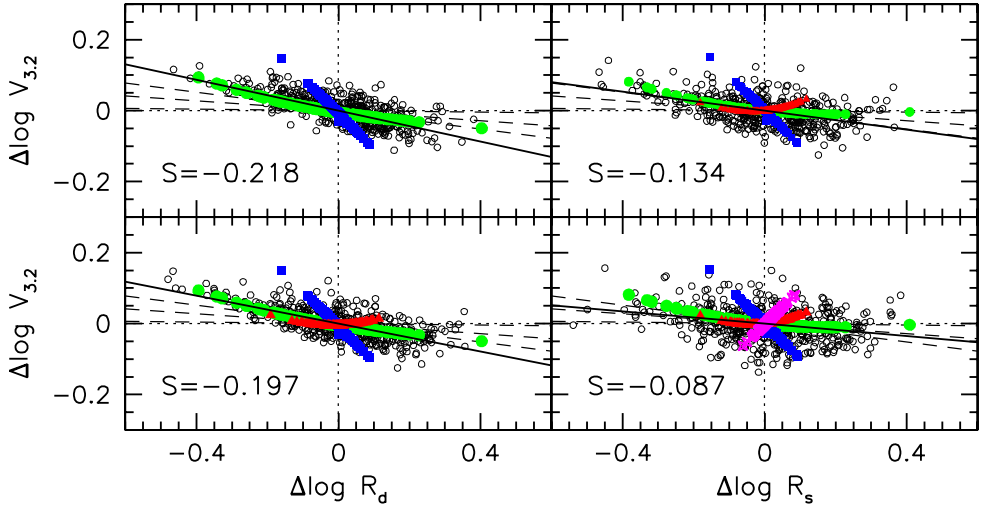}{2.3in}{0}{70}{70}{-220}{-280}
\caption{Correlations between residuals of VL and RL at a
  given L for our fiducial model  with no AC in a halo of $\Vvir=100$.
  Top-left:  scatter  in  $\lambda$  (green  circles)  and  $c$  (blue
  squares) only.  Bottom-left: added scatter in $\md$ (red triangles).
  Top-right:   added    density   threshold   for    star   formation.
  Bottom-right: added scatter in  $\YI$ (magenta stars).  The 4 panels
  refer  to  residuals  at  a  given $\Md$,  $\Md$,  $\Ms$  and  $\Ms$
  respectively.  The linear fit  ($\Delta\log V$ on $\Delta\log R$) is
  shown (solid line) and its  slope is marked.  The mean and 2$\sigma$
  scatter about the observed relation are shown (dashed lines).}
\end{figure}

As noticed  by CR99, the degree  of correlation between the  VL and RL
residuals  at a given  $L$ provides  a powerful  constraint.  Courteau
\etal (2003) measured a weak anti-correlation between these residuals,
fit  in   log  space   by  a  line   of  slope   $S\equiv  {\Delta\log
V}/{\Delta\log  R}=-0.07\pm0.03$, confirming  the  original result  of
CR99. Note that $S=0$ for completely uncorrelated residuals.

The slope $S$  is connected to the relative  contributions of the disc
and halo to the potential,  and hence to the circular velocity $\VII$,
measured at $\RII$.  The  disc contributes $\Vdisc^2 \propto \Ms/\Rs$,
namely  $S=-0.5$.   A  fixed   halo,  not  reacting  to  the  baryonic
contraction, contributes  $\Vhalo^2 \propto \Rs$,  and thus ($S=+0.5$)
for  an  NFW density  cusp.   One can  therefore  expect  that a  weak
anti-correlation,  as  observed,  would  be obtained  with  comparable
contributions of disc and halo.  AC increases the halo contribution to
$\VIII$  in a  way that  anti-correlates  with the  disc radius,  thus
lowering $S$ toward small or  negative values.  In order to maintain a
small negative $S$  with AC, the contribution of  the halo, therefore,
has  to  be  larger.   Indeed,  assuming AC,  CR99  interpreted  their
uncorrelated  residuals  as  indicating  a  ``sub-maximal"  disc  with
$\Vdisc/\VII  \simeq 0.55$.   Matching the  low negative  $S$ observed
with  our  models  that  include  AC,  requires  very  low  values  of
$\mdbar/\lambar$,  pushing  the  parameters outside  their  reasonable
ranges.   Furthermore,  this  is  at  the expense  of  increasing  the
zero-point  offsets  (\S  3.2),  so  this model  does  not  provide  a
simultaneous fit to the residuals {\it and} the zero-points.
 
When the AC is eliminated,  our model obeys the zero-point constraints
and   reproduces  the   weakly   anti-correlated  residuals.    Fig.~2
illustrates how the various sources of scatter affect the distribution
of residuals $\Delta\log  V$ and $\Delta\log R$ at  a given $M$.  When
we use  the {\it total\,} disc  mass and radius, with  scatter only in
$\lambda$ and $c$, we still find a strong anti-correlation, $S=-0.22$.
Adding  scatter in $\md$  reduces it  to $S=-0.20$.   When considering
{\it  stellar\,} mass  and radius,  we obtain  a  significantly weaker
$S=-0.13$; this is because  $\Sigmacrit$ links $\Ms$ to $\lambda$ (van
den Bosch 2000, Firmani \& Avila-Reese 2000).  When adding the scatter
in $\YI$, there is another significant reduction to $S=-0.087$, due to
the  fact  that  the  $\YI$  scatter  by  itself  induces  a  positive
correlation of $S=+1$ (bottom-right  panel).  Finally, when adding the
(uncorrelated) measurement errors  we obtain the residual distribution
shown  in Fig.~3,  with  $S=-0.074$  --- in  good  agreement with  the
observations.

\begin{figure}[t]
\plotfiddle{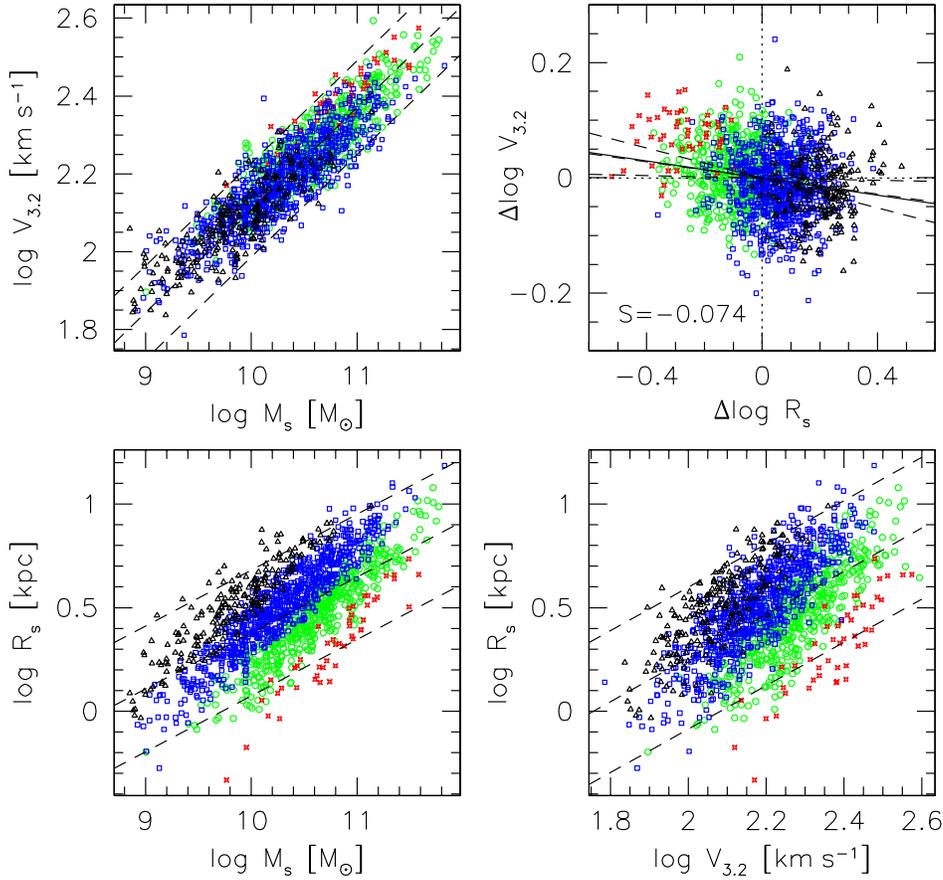}{4.2in}{0}{65}{65}{-195}{-113}
\caption{Distribution of structure parameters for our fiducial model 
  (no  AC, Kroupa IMF)  versus the  observations. The  top-right panel
  refers to  the residuals.  Colored symbols refer  to central surface
  density $[\Msol\,pc^{-2}]$: $\Sigma_0 < 200$ (black triangles); $200
  < \Sigma_0  < 630 $ (blue  squares); $630 < \Sigma_0  < 2000$ (green
  circles); and $2000  < \Sigma_0$ (red stars). Dashed  lines refer to
  the mean and 2$\sigma$ scatter in the data.}
\end{figure}

Fig.~3 shows  the joint distribution  of structural parameters  in our
successful model compared to  the observed results.  The model assumes
no  adiabatic contraction.   It  adopts the  $c$  distribution of  CDM
haloes.  It assumes $\lambar = 0.035$, as for haloes, but $\sigma_{\ln
\lambda}=0.25$, about half the width of the halo distribution.  It has
$\mdbar   =0.07(\Mvir/10^{12}    \Msun)^{0.16}$,   with   $\sigma_{\ln
{\md}}=0.23$.  The  translation to  $\Ms$ is done  using a  $\YI$ that
corresponds to a Kroupa  IMF, with $\sigma_{\YI}=0.23$.  A measurement
error of $\sigma_{\ln}=0.09$ is assigned to each of the three observed
quantities.

\section{Conclusion}

We used the  slope, zero-point, scatter and residuals  of the observed
velocity-mass and radius-mass distributions for disc galaxies to place
constraints  on the  basic CDM-based  model of  galaxy  formation.  By
applying these constraints {\it simultaneously}, we obtain non-trivial
conclusions.    Our  simple   model  assumes   independent  log-normal
distributions  for  3  random   variables  at  each  halo  mass:  halo
concentration  $c$, baryonic  spin $\lambda$,  and disc  mass fraction
$\md$.   Additional   sources  of   scatter  come  from   the  stellar
mass-to-light ratio $\YI$ and the measurement errors.

Contrary to previous  studies, we take account of  the fact that discs
are not  purely stellar, and  that the stellar mass-to-light  ratio is
not  constant. For our  model-data comparison,  we have  converted the
observed $I$-band  luminosities to stellar masses,  using the relation
between color and $\YI$.   This immediately demonstrates that even the
slopes of the scaling  relations are not automatically reproduced from
the  basic  halo  virial   relation  $\Mvir  \propto  \Vvir^3  \propto
\Rvir^3$.   We find  that for  a  simultaneous match  of the  observed
slopes, given the $c$ distribution  as predicted for CDM haloes, stars
have to form  only above a threshold density  $\Sigma_{\rm crit}$, and
the stellar  mass fraction  has to increase  with halo mass.   This is
consistent with  the expected effects of supernova  feedback, known to
be more  efficient in  suppressing star formation  in haloes  of lower
masses (e.g.,  Dekel \& Silk  1986; van den  Bosch 2002; Dekel  \& Woo
2003).

The  RM  scatter  is  dominated  by  the  scatter  in  spin  parameter
$\lambda$.   The relatively small  scatter observed  requires robustly
that  $\sigma_{\rm\,ln  \lambda}  \simeq0.25$,  about half  the  value
predicted for  typical CDM haloes.  This  could be due  to a selection
effect in the data (e.g. against  LSB or S0 galaxies), or may indicate
that  discs  form  in a  special  subset  of  haloes with  a  narrower
$\lambda$  distribution.   Interestingly,   haloes  with  a  quiescent
history, which are  the natural hosts of discs, indeed  tend to have a
narrower   $\lambda$   distribution   (D'Onghia  \&   Burkert   2004).
Alternatively  it may indicate  that the  baryons somehow  developed a
narrower $\lambda$  distribution than that of the  dark matter haloes.
The VM relation  is much tighter because it  is largely independent of
the scatter in  $\lambda$.  The intrinsic VM scatter  is determined by
the scatter  in both $c$ {\it  and} $\YI$, as each  alone violates the
constraint from the uncorrelated residuals.

Matching the  zero-points of the scaling relations,  combined with the
weak  anti-correlations between  the residuals,  turns out  to  be the
greatest  challenge.   It  could  have been  achieved  by  drastically
lowering  $c$ relative  to  the predictions  for typical  $\Lambda$CDM
haloes, or  by assuming an  unrealistic top-heavy IMF.  In  both cases
the parameters  have to be  pushed away from their  reasonable limits,
and the  zero-points are matched without  reproducing the uncorrelated
residuals.   One could  alternatively  achieve weakly  anti-correlated
residuals by  assuming unrealistically low $\md/\lambda$,  but only at
the expense of worsening the zero-point offsets.

The only way we found  to match simultaneously the zero-points and the
uncorrelated residuals is by eliminating the adiabatic response of the
dark  matter to the  condensation of  the gas  into the  central disc.
Once the AC is removed, it becomes relatively straightforward to match
simultaneously  all the  constraints  (though with  a  low scatter  in
$\lambda$).   Except  for a  model  with  unrealistically low  stellar
mass-to-light ratios, no other  reasonable fix of the model parameters
comes even close to  simultaneously matching the zero-points {\it and}
the  residual constraints,  making our  conclusion quite  robust.  Our
successful model with no AC  predicts that discs contribute about 50\%
to $\VII$ in  the smaller, low surface brightness  (LSB) galaxies, but
as  much  as 80\%  in  the  brighter,  high surface  brightness  (HSB)
galaxies.   Thus, whereas the  LSBs are  indeed sub-maximal,  with the
haloes dominating  the inner regions,  the HSBs are closer  to maximum
discs.

In an ongoing  work, we attempt to quantify  the degree of contraction
permitted by the data; yet to  be confirmed is our impression that the
corrected AC suggested by Gnedin \etal (2004) is not enough.  Once the
discs  are  not  built  by  smooth  spherical  infall,  the  adiabatic
contraction could,  in principle, be counter-balanced by  a variety of
processes, and  may even end up  as an expansion. One  such process is
dynamical  friction (e.g.   El-Zant \etal  2004; Ma  \& Boylan-Kolchin
2004), which could  be important if the discs  are built by relatively
big clumps (e.g. Maller \&  Bullock 2004).  Indeed, the physics of gas
infall  in dark haloes  suggests that  most big  discs have  formed by
clumpy,  cold streams  rather than  smooth infall  (Birnboim  \& Dekel
2003; Dekel \& Birnboim 2005).

Two caveats are worth mentioning.   First, we should verify using more
accurate color information that  there is no systematic uncertainty in
the conversion of $\LI$ to $\Ms$, which could cause an overestimate of
$\YI$ for a  given IMF.  Second, our analysis  assumed independence of
the scatter in  the variables at a given $M$. We  suspect that $c$ and
$\lambda$  may be  slightly correlated,  e.g. via  the  halo accretion
history (Wechsler  \etal 2002), and one could  imagine scenarios where
$\md$  is also  slightly correlated  with the  others. To  what extent
these weak correlations may affect our conclusions remains to be seen.

\smallskip

We thank A. Maller \&  S.M. Faber for stimulating discussions.  AD has
been  partly supported  by ISF  213/02, NASA  ATP NAG5-8218,  a Miller
Professorship at UC Berkeley,  and a Blaise Pascal International Chair
in  Paris. SC  recognizes the  support  of NSERC  through a  Discovery
grant.

\end{document}